\documentclass[aps,preprint]{revtex4}%
\usepackage{amsfonts}
\usepackage{amsmath}
\usepackage{amssymb}
\usepackage{graphicx}%
\begin{document}
\preprint{HEP/123-qed}
\title[Short title for running header]{Magnetoresistance of p-GaAs/AlGaAs structures
in the vicinity of metal-insulator transition: Effect of
superconducting leads. }
\author{N.V.Agrinskaya, V.I.Kozub, A.V.Chernyaev, D.V. Shamshur, A.A. Zuzin. }
\affiliation{A.F.Ioffe Physico-Technical Institute}
\pacs{72.20 Ee, 74.45+c, 74.50+r}

\begin{abstract}
Experimental and theoretical studies on transport in semiconductor
samples with superconducting electrodes are reported. We focus on
the samples close to metal-insulator transition. In metallic
samples,  a peak of negative magnetoresistance at fields lower
than critical magnetic field of the leads was observed. This peak
is attributed to restoration of a single-particle tunneling
emerging with suppression of superconductivity. The experimental
results allow us to estimate  tunneling transparency of the
boundary between superconductor and metal. In contrast, for the
insulating samples no such a peak was observed. We explain this
behavior as related to properties of transport through the contact
between  superconductor and hopping conductor. This effect can be
used to discriminate between weak localization and strong
localization regimes.

\end{abstract}
\volumeyear{year}
\volumenumber{number}
\issuenumber{number}
\eid{identifier}
\date[Date text]{date}
\received[Received text]{date}

\revised[Revised text]{date}

\accepted[Accepted text]{date}

\published[Published text]{date}

\startpage{1}
\endpage{10}
\maketitle


Recently  we reported an observation of crossover from strong to
weak localization in 2D p-GaAs/AlGaAs structures \cite{Agrins1}.
The magnetoresistance of our samples in weak localization regime
has demonstrated, in addition to the standard antilocalization
behavior, a small peak of negative magnetoresistance (NMR) at weak
fields $< 0,02$ T. The nature of this peak was not clear. It could
not be attributed to weak localization since at higher magnetic
fields the samples clearly demonstrated positive magnetoresistance
related to antilocalization. Since the peak disappeared at the
critical temperature of In contacts,   $3.4 $ K, it was natural to
assume that it originates from the superconducting contacts.
However, simple considerations would predict PMR at magnetic field
destroying superconductivity of the contacts. Another point was
that the effect was not observed for the hopping regime even when
the resistance of the samples was not much larger than the
resistance of the metal samples. Thus, further analysis of this
effect was necessary.

Usage of superconducting contacts is a common practise in studies
of the samples close to the metal-insulator transition (MIT) or
deep in the hopping regime. However, though a significant
attention was paid to the properties of a superconductor - normal
metal interface, we are not aware of  studies of the interface
between superconductor and the sample close to MIT transition.

Here we present results of experimental and theoretical studies of
a role of superconducting contacts to structures in the vicinity
of MIT from both sides of the transition. We will prove that the
peak of the magnetoresistance mentioned above is related to a
presence of an insulating barrier between the superconductor and
semiconductor. This barrier suppresses the Andreev reflections,
the single-particle channel being affected by the superconducting
gap. Thus the mechanism of the observed negative magnetoresistance
is suppression of the gap by the magnetic field. We will show that
this model allows to explain the experimental results in detail.

The samples on the dielectric side of MIT do not demonstrate any
traces of the effect - even with an account of the sample
resistance increase. We will show that this fact is related to
specific properties of the hopping transport including a larger
value of the effective energy band than for metals and the
topology of the percolation cluster. We believe that the unusual
magnetic field dependence of electron transport in systems with
superconducting electrodes can be used as a tool  to discriminate
between the regimes of weak and strong localization.

\section{Experiment}

We have chosen $\mathrm{GaAs/Al}_{0.3}\mathrm{Ga}_{0.7}\mathrm{As}
$ multi-well structures  with the well widths ($d$) of 10nm, and
some larger barrier width of 25nm, doped by an acceptor dopant,
Be. The binding energy  of Be dopant is $E_{A}\sim 28 $ meV which
yields the localization length $a_{b}=2$ nm being much less than
$d$). The method of growing multilayer structures by molecular
beam epitaxy was described in our work \cite{Agrins2}. In sample
1, by  selective doping of the central regions of the wells with
relative widths 1/3 we prepared the system where the lower Hubbard
(LH) impurity band was formed. In the sample 2 by  selective
doping of the central regions of both wells and barriers (with
equal doping concentrations) we prepared a system where the upper
Hubbard (UH) impurity band was partly occupied in the equilibrium.
The dopant concentration $N_{a} \sim(1-2)\cdot10^{12}$ cm$^{-3}$
was near the critical concentration for the metal-insulator
transition in 2D structures, $N_{c}^{1/2} a_{b} \sim 0,3$. The
contacts were produced by firing indium with a low zinc
concentration during 2 min at  temperature 450 C
${{}^\circ}$%
.

Close to the room temperature, the temperature dependences of hole
concentration show activation behavior caused by the transition of
holes from the impurity band to valence band. From these parts of
the curves we estimated the Fermi energies in samples 1 and 2 -
15-20 meV and 6 meV, respectively.  In the sample 1 Fermi level is
located in LH band, while in the sample 2 - in the UH band. At low
temperatures conductivity of these two samples depends on
temperature very weakly (fig.1). As we have shown, it can be
described by the weak localization theory \cite{Agrins1}.

A standard antilocalization behavior (positive magnetoresistance)
was observed in our samples for magnetic fields $> 0.02$T. For
lower fields these two samples demonstrated also a small region of
negative magnetoresistance (NMR) (fig.2). Such a behavior was
observed only for temperatures lower than $\sim 3.4 $K which
corresponds to the temperature of the  superconducting transition
in In. The  NMR magnitude,  $(1- 10)\%$, depends on a concrete
realization of the contact. Shown in Fig.3 is the temperature
dependence of this low field NMR in the temperature region 3-1.2
K. One can see the increase of this NMR magnitude with a
temperature decrease following by saturation at temperatures 1.2 -
0.6 K.

\section{Theory}

Let us consider a tunnel contact between a superconductor and a
semiconductor sample.  First we will discuss the situation on the
metal side of the MIT, that is of the tunneling between a
superconductor and a normal metal. As it is known, see, e. g.,
\cite{tunnel}, the single-electron tunneling current can be
written as:
\begin{equation}
I=\frac{4\pi e}{\hbar}\left\vert T_{0}\right\vert ^{2}\int
d\varepsilon \, \nu _{1}(\varepsilon+eV)\, \nu_{2}(\varepsilon)\,
\left[ n_{1}(\varepsilon )-n_{2}(\varepsilon+eV)\right]\, .
\end{equation}
where $T_{0}$ is the tunneling matrix element, $\nu_{n}$ is the
density of states in the 2D metal. The density of quasiparticle
states in a superconductor is
\begin{align}
\nu_{s}  &  =\nu_{n2}\frac{\left\vert \varepsilon\right\vert
}{\sqrt {\varepsilon^{2}-\Delta^{2}}}\quad \text{at} \quad
\left\vert \varepsilon\right\vert
>\Delta \, ;\nonumber\\
\nu_{s}  &  =0 \quad \text{at} \quad \left\vert
\varepsilon\right\vert <\Delta \, .
\end{align}
Correspondingly, one obtains
\begin{equation}
I_{N-S}(V)=\frac{4\pi}{\hbar}e\left\vert T_{0}\right\vert ^{2}\nu_{n1}\nu_{n2}%
{\textstyle\int\limits_{\left\vert \varepsilon\right\vert >\Delta}}
\left[  n_{1}(\varepsilon)-n_{2}(\varepsilon+eV)\right]  \frac{\left\vert
\varepsilon\right\vert d\varepsilon}{\sqrt{\varepsilon^{2}-\Delta^{2}}}
\label{2}
\end{equation}
At low temperature,  $T<<\Delta$, the integral (\ref{2}) in the
linear approximation in $V$ can be estimated as
\begin{equation}
I_{N-S}(V)=VG\sqrt{\frac{2\pi\Delta}{T}}e^{-\Delta/T} \label{low}%
\end{equation}
where
\[
G=\frac{4\pi}{\hbar}e^{2}\left\vert T_{0}\right\vert ^{2}\nu_{n1}\nu_{n2}%
\]
is the contact conductance when the superconductor is in the
normal state.

Now let us consider the behavior near the critical temperature
$T\rightarrow T_{c}$. In the linear regime,  $eV<T$, the direct
estimate of Eq. (\ref{2}) yields
\begin{equation}
I_{N-S}(V)=VG\left[  1-\left(  \frac{\Delta}{2T}\right)  ^{2}\right]
\label{met}
\end{equation}
This quasiparticle current vanishes at $T \rightarrow 0$. In
addition, there is also a contribution of the Andreev reflections
which does not vanish,  (see e.g. \cite{tunnel}). Let us estimate
the temperature at which the single-particle contribution crosses
over to the Andreev contribution. The latter can be estimated as
\begin{equation}
G_{A}=\frac{e^{2}}{\hbar}A\Gamma^{2}%
\end{equation}
where $\Gamma = gN\left\vert T_{0}\right\vert ^{2}S$ is the
tunneling transparency and A is a constant. The quasiparticle
contribution can be rewritten as
\begin{equation}
G_{T}=\frac{e^{2}}{\hbar}\sqrt{\frac{2\pi\Delta}{T}}e^{-\Delta/T}A\Gamma
.
\end{equation}
Correspondingly, these contributions are equal at some crossover
temperature $T^{^*}$. Considering $T^{*}$ as given, one can
estimate the tunneling transparency $\Gamma$ in a simple way:
\begin{equation}
\Gamma=\sqrt{\frac{2\pi\Delta}{T^{^{\prime}}}}e^{-\Delta/T^{^{\prime}}%
}\label{cross}
\end{equation}
Now let us consider the insulator side of the MIT transition when
one has tunneling between the superconductor and semiconductor in
hopping regime. In this case the single-particle tunneling between
semiconductor and superconductor banks can be controlled either by
the direct resonant electron tunneling or by the phonon-assisted
tunneling. The latter process dominates if the effective hopping
energy band, $\varepsilon_{0}$, is less than the superconducting
gap. In this case the temperature behavior of the conductance is
given as
\begin{equation}
G\propto e^{-\frac{\left(  \Delta-\epsilon_{0}\right)  }{T}}\label{9}
\end{equation}
To the contrary, if $\varepsilon_{0}>\Delta$ the contribution of resonant
tunneling is expected to dominate.

The character of transport is also expected to be sensitive to the
strength of the tunneling barrier. Indeed, transport in the
semiconductor is controlled by the percolation cluster which
allows self-averaging of the conductivity. Thus the contact
resistance is the resistance between the superconductor and the
percolation cluster. It consists of a sum of the resistance
related to the last hop between some localized state in the
semiconductor and the superconductor, and the resistance of the
branch connecting this localized state to the percolation cluster.
If the tunnel barrier transparency $\Gamma$ is much less that the
typical hopping exponent $\exp (-2r_{h}/a) $, where  $r_{h}$ is
the typical hopping length, than the contact resistance is
dominated by the "last" hop. In its turn, the contact conductance
is a sum of the conductances corresponding to these hops. This
fact allows us to average over these conductances (see
\cite{Shklovskii}. For each of the localized state $i$ the
corresponding conductance is
\begin{equation}
G_{T,i}\propto\frac{e^{2}}{\hbar}N\left\vert T_{0}\right\vert ^{2}\frac{1}%
{aT}\frac{\varepsilon_{i}}{\sqrt{\varepsilon_{i}^{2}-\Delta^{2}}}\exp\left(
-\frac{2x_{i}}{a}\right)
\end{equation}
where $x$ is the coordinate normal to the contact. Correspondingly, for the
average one has
\begin{equation}
G_{SS}\propto\frac{e^{2}}{\hbar}\left\vert T_{0}\right\vert ^{2}\frac{N}%
{aT}{\int\limits_{\Delta}^{\epsilon_{0\text{ }}}}d\varepsilon \,
g(\varepsilon)\frac{\varepsilon}{\sqrt{\varepsilon^{2}-\Delta^{2}}%
}{\int\limits_{\text{ }}}d^{2}r\, dx\exp\left(
-\frac{2x}{a}\right)
\label{7}%
\end{equation}
where $g$ , $N$ is the densities of localized states in
semiconductor and metal respectively. Evaluating the expression
(\ref{7}) one obtains
\begin{equation}
G_{SS}\propto\frac{e^{2}}{\hbar}\Gamma\frac{\varepsilon_{0}}{T}\sqrt{1-\left(
\frac{\Delta}{\varepsilon_{0}}\right)  ^{2}} \label{GSS}
\end{equation}

Now let us consider the situation when $\Gamma>\exp (-2r_{h}/a)$.
In this case the contact between the superconductor and the
percolation cluster is supported by some branches of hopping
resistors. Since these branches are in parallel, the conductance
of this structure is given as
\begin{equation}
G_{SS}=%
{\textstyle\sum\limits_{i}}
\frac{1}{  R_{T,i}+R_{c,i}  } \label{c}%
\end{equation}
where $R_{T,i}=G_{T,i}^{-1}$ is the resistance of the hop form the
last localized state to superconductor, while $R_{c,i}$ is the
resistance of the branch connecting this last hop with the
percolation cluster. One can separate the contact contribution to
resistance as
\begin{equation}
\left(  \sum_{i}\frac{1}{(R_{T,i}+R_{c,i})}\right)  ^{-1}-\left(  \sum
_{i}\frac{1}{R_{c,i}}\right)  ^{-1}=\frac{\displaystyle{\sum_{i}\frac{R_{T,i}}%
{R_{c,i}(R_{T,i}+R_{c,i})}}}{\displaystyle{\sum_{i}\frac{1}{R_{c,i}}\sum_{i}\frac{1}%
{R_{c,i}+R_{T,i}}}}\, .
\end{equation}
Since the branches support the current flow through the system,
according to ideas of the percolation theory they should also be
considered as a part of the percolation cluster. Thus one expects
that all of $R_{c,i}$ are of the order of that corresponding to
the percolation threshold, $R_{c}$. So the only average should be
taken with respect to the localized state $i$ corresponding to the
last hop, actually - with respect to $x_{i}$. One notes that the
upper limit for $x_{i}$ is given by some critical value,  $x_{i}$,
corresponding to $R_{T,i}=R_{c}$. The larger resistances do not
enter the percolation cluster. Then, one notes that for
$x_{i}<x_{c}-a$ one has $R_{T,i}<<R_{c}$ and the corresponding
paths do not contribute effectively to the contact resistance.
Thus only a small part of the branches given by a ratio
$a/x_{c}\sim1/\xi$ is important for the contact magnetoresistance.

Combining this estimate with the one given by Eq. (\ref{GSS}) one
notes that for the hopping conductivity the contact
magnetoresistance is suppressed with respect to the metal
conductor due to the  two issues: (i) since the hopping energy
band width $\varepsilon_{0}$ entering Eq. (\ref{GSS}) is larger
than $T$ entering Eq. (\ref{met}) and additional factor
$(T/\varepsilon_{0})^{2} = \xi^{-2}$ appears at $\Delta<
\varepsilon_{0}$; (ii) in contrast to a metal, in the hopping case
only small part of tunneling events contributes to the contact
magnetoresistance which gives a factor $1/\xi$. As a result, the
contact magnetoresistance in hopping regime is suppressed with
respect to metal by a factor $\xi^{-3}$.

Actually the measurable quantity is a sum of the the contact
resistance and the sample resistance. So an increase of the sample
resistance in the hopping regime $\propto\exp\xi$ means a decrease
of \textit{relative} magnetoresistance if the latter originates
from the contact contribution. As a result, a suppression of the
\textit{relative} magnetoresistance in the hopping regime is given
by the factor
\begin{equation}
\label{hop}\xi^{-3}\exp- \xi
\end{equation}
rather than by a factor $\exp (- \xi)$ as one may conclude from
naive considerations.

Note that as concerns the hopping regime we restricted ourselves
by single particle tunneling. According to Eq. (\ref{9}), this
channel is exponentially frozen out at $T \rightarrow0$. This
conclusion still holds if the tunnel barrier is weak or even
absent since it is based on the  single-particle character of
transport impossible in a superconductor at $T = 0$. As well known
for the interface between a superconductor and a normal metal it
is the two-particle Andreev reflections that are responsible for
the low temperature transport. However the typical processes
leading to hopping transport are single-particle ones. Note that
the activation exponential factor $\exp (-\Delta /T)$ ) at small
temperatures can be much smaller than variable range hopping
factor $\exp (- \xi) $. This is the case e.g. for studies
\cite{Rentzch} of hopping transport in CdTe structures with In
contacts where the temperatures were as low as $\sim30 $mK.  Thus
a process similar to the Andreev reflection, i. e involving
tunneling of Cooper pairs, should be considered.  To the best of
our knowledge, no detailed studies of such transport were reported
until now. We are going to address this topic in a special paper.

\section{Discussion}

As clearly seen, the NMR at low fields can be only related to
presence of a superconductor since the effect is absent at $T
> T_{c}$. In principle, the magnetoresistance could be also related to
interference contribution to Andreev tunneling, see e. g.
\cite{Nazarov}. However, our samples being  close to MIT
correspond to  a limit of extreme dirty metal and thus
characteristic magnetic field scales for the interference effects
are much larger than critical magnetic field for In (see e. g. our
studies \cite{Agrins1}) . At the same time, the observed peak
occurs at $ H \lesssim 0.1$ T, that is of the order of the
critical field for In but much less than required for the
interference effects.

In Fig. 3 we have plotted the magnitude of the magnetoresistance
peak as a function of temperature for the sample on the metal side
of MIT. As it is seen, it is proportional to $\propto(T_{c} - T)$
in the vicinity of $T_{c}$. That agrees with Eq. (\ref{met}),
$\Delta R \propto\Delta^{2}$ since according to the BCS theory
$\Delta\propto(T_{c} - T)^{1/2}$.

For lower temperatures Eq. (\ref{low}) predicts exponential
increase of magnetoresistance which can be seen from Fig. 3 . With
a further decrease of temperature the magnetoresistance saturates
due to a bypassing of the Andreev channel. According to  Eq.
\ref{cross}, the crossover temperature allows us to estimate the
tunnel barrier transparency. Making use of the experimental data
one concludes that $T^{^{*}} \approx 1.1$ K.  It is well known
that the superconducting transition temperature for In is
$T_{c}=3.4$ K and thus $\Delta\approx 6 $K. Substituting $\Delta$
and $T^{^{*}}$ into Eq. (\ref{cross}) we estimate the tunneling
probability as $\Gamma \approx 0.05$.

On the Fig. 3 we have also plotted theoretical curve resulting
from the summation of the single-particle contribution calculated
above and the contribution of the Andreev tunneling. Unfortunately
the proper analysis of this latter channel at temperatures close
to $T_c$ is difficult due to the fact that our "normal metal" is
actually a dirty semiconductor and is characterized by rather
small spin-orbital scattering times ($> 10^{-11}s$) and phase
relaxation times ($\sim10^{-11}s$) \cite{Agrins1}. To the best of
our knowledge, the detailed theory of Andreev reflections in the
vicinity of $T_{c}$ with an account of the factors mentioned above
is absent. We have concluded that the best fitting of the
experimental data is obtained if we approximate the Andreev
channel contribution at $T \sim T_c$ as $\propto
(\Delta(T))^2/(\Delta(0))^2$.

Now let us turn to the magnetoresistance in the hopping regime. As
follows from Fig. 2, in the hopping regime there is no  trace of
the magnetoresistance peak at $T=1.4$ K with an accuracy at least
of the order of $0.01\%$. Note that the corresponding magnitude of
the relative magnetoresistance peak in the metal sample was of the
order of $\sim 1\%$, Fig. 2  . Such a suppression of the relative
magnetoresistance in the hopping regime can hardly be explained by
a simple increase of resistance. Indeed, the latter is larger than
that of the metal sample only by a factor $<100$. However, this
behavior is easily explained by our Eq. (\ref{hop}) predicting
much stronger suppression of the magnetoresistance than following
from the resistance ratio. In our opinion, this effect can be used
to discriminate between the regimes of weak  and localization even
in the crossover region between the two regimes. The reason is in
the  principal differences in the physical picture of transport
between the regimes which are not clearly seen in the value of the
resistance itself.

This is why we believe that  our model is adequate for the
experimental findings.

To conclude, we studied magnetoresistance of 2D p-GaAs/AlGaAs
structures with superconducting electrodes (In) close to the
metal-insulator transition. We demonstrated that the observed weak
field magnetoresistance peak observed  for metallic samples is due
to  restoration of the single-particle tunneling through the
superconductor-semiconductor boundary with suppression of
superconductivity. The crossover between the two tunneling regimes
allowed us to estimate the tunneling transparency. The samples on
the dielectric side of MIT did not demonstrate such a peak. the
suppression of the magnetoresistance peak in the hopping regime is
confirmed by theoretical analysis. We suggest to use this effect
for discrimination between the regimes of weak and strong
localization.

\section{Acknowledgements}

We are grateful to V. M. Ustinov and his group for growing the
corresponding structures and to A. L. Shelankov for discussions.
We are also indebted to Yu. M. Galperin for reading the manuscript
and many valuable remarks. This work was supported by Russian
Foundation for Basic Research (project no. 03-02-17516 ) and by
the Ministry of Science (project no. 9S19.2). A. A. Z. is grateful
to International Center for Fundamental Physics in Moscow and to
the Fund of non-commercial programs ``Dynasty".

\vfill\eject

Figure captions.

Fig.1 Temperature dependences of the conductivity for 3 samples:
samples 1,2 are in WL regime, sample 3 - in SL regime. Sample 1 -
wells (10 nm) and barriers are doped ( bulk Be concentration is
$6\cdot 10^{17} cm^{-3}$), sample 2 - only wells (15 nm) are doped
( bulk Be concentration is $10^{18} cm^{-3}$), 3 - wells (15 nm)
and barriers doped ( bulk Be concentration is $5\cdot 10^{17}
cm^{-3}$).

Fig.2 Low field magnetoresistance at different temperatures: a)
for sample 1 (WL regime), b) for sample 3 (SL regime).

Fig.3 Temperature dependence of NMR peak magnitude for sample 1
(experiment and theory).

Theoretical fitting equation $\frac{\Delta R}{R_{0}}=\left(  \frac{1}%
{\frac{R_{N}}{R_{T}}+\left(  \frac{\Delta(T)}{\Delta(0)}\right)  ^{2}%
\frac{R_{N}}{R_{A}}}-1\right)  \frac{R_{N}}{R_{0}}$.
 Where $R_{T}$
,$R_{A}$, $R_{N}$ - are the single-particle tunneling , Andreev
and normal state contact resistances respectively.

The fitting parameters are $\frac{R_{N}}{R_{0}}\approx4.2$ , $\frac{R_{N}%
}{R_{A}}\approx0.33$

\bigskip


\begin{thebibliography}{9}
\bibitem {Agrins1}N. V. Agrinskaya, V. I. Kozub, D. V. Poloskin, A. V. Chernyaev, D. V.
Shamshur, JETP Letters \textbf{80}, 30 (2004).

\bibitem {Agrins2}N. V. Agrinskaya, V. I. Kozub, Yu. L. Ivanov et. al., JETP
\textbf{93}, 424 (2001).

\bibitem {tunnel}\textit{Tunneling phenomena in solids}, ed. E. Burstein and S. Lundvist
(Plenum, New York, 1969).

\bibitem {Rentzch}N. V. Agrinskaya, V. I. Kozub, R. Rentzsch, JETP \textbf{84}, 814
(1997).
\bibitem{Shklovskii} A. I. Larkin, B. I. Shklovskii. Phys. Stat. Sol.
B, {\bf 230}, 189 (2002).

\bibitem {Nazarov}F.W.J.Hekking, Y.V.Nazarov. Phys. Rev. B, \textbf{49}, 6847 (1994).
\end{thebibliography}
\end{document}